\documentclass[
amsmath, aps, prl, %
twocolumn
]{revtex4}
\usepackage{bm}
\usepackage{graphicx}
\usepackage{amssymb}

\begin{document}
\title{Stretching  vortices as a basis for the theory of turbulence}
\author{V.A. Sirota\footnote{Electronic addresses: sirota@lpi.ru, zybin@lpi.ru}
and K.P. Zybin$^*$}
\affiliation 
{P.N.Lebedev Physical Institute of RAS, 119991, Leninskij pr.53,
Moscow, Russia }


\begin{abstract}

 Turbulent flows play an important role in many
aspects of nature and technics from sea storms to transport of
particles or chemicals. Transport of energy from large scales to
small fluctuations is the essential feature of three-dimensional
turbulence. What mechanism is responsible for this transport and how
do the small fluctuations appear? The conventional conception
implies a cascade of breaking vortices. But it faces crucial
problems in explaining the mechanism of the breaking, and fails to
explain the observed long-living structures in turbulent flows.

We suggest a new concept based on recent analysis of stochastic
Navier-Stokes equation: stretching of vortices instead of their
breaking may be the main mechanism of turbulence. This conception is
free of the disadvantages of the cascade paradigm; it also does not
need finite-time singularities to explain the observed statistical
properties of turbulent flows. Moreover, the introduction of the new
conception allows immediately to get velocity scaling parameters
well consistent with experimental data.
\end{abstract}


\maketitle

Turbulent motion implies the presence of multiple vortices of
different scales. The general physical conception that has been
associated with turbulence since the beginning of the XX century is
the conception  of cascade of breaking vortices introduced by
Richardson. Just as one rock falling down breaks to two pieces, then
to four pieces, etc., making a rockslide, or as energetic cosmic
particle produces an avalanche of elementary particles after several
generations of scatterings, large eddies in turbulent flow are
believed to break consequently into smaller vortices, producing
turbulence (Fig. 1).  In each of these processes there is a physical
quantity that conserves: in the case of crushing rocks it is the
total mass of each generation, in the case of cosmic particles it is
the total energy. In turbulent cascade, the  conserving quantity is
the energy flux from larger to smaller scales. Actually, energy is
passed to a flow at large scales (from moving boat in the lake, or
rotating turbine in washing machine).  The only scale where it is
lost is small dissipation scale which is determined by viscosity of
the liquid. This is 'the end of the avalanche', turbulence ceases at
this scale. Intermediate generations of vortices belong to so-called
inertial range of scales, they conduct energy from larger to smaller
vortices (Fig. 2).

\begin{figure}[h]
\vspace*{-0.4cm} \hspace*{-0.5cm}
\includegraphics[width=8cm]{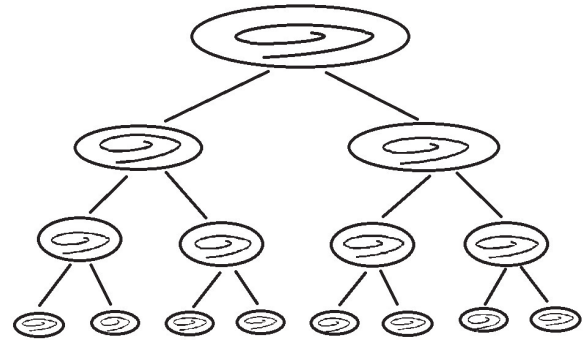}   
  \vspace*{-0.6cm}
\caption{Cascade of breaking vortices: the well-known scheme
associated with Kolmogorov's theory (see, e.g., \cite{Frisch}) but
incompatible with existence of long-living structures. }
\end{figure}

\begin{figure}[h]
\vspace*{-0.4cm} \hspace*{-0.5cm}
\includegraphics[width=8cm]{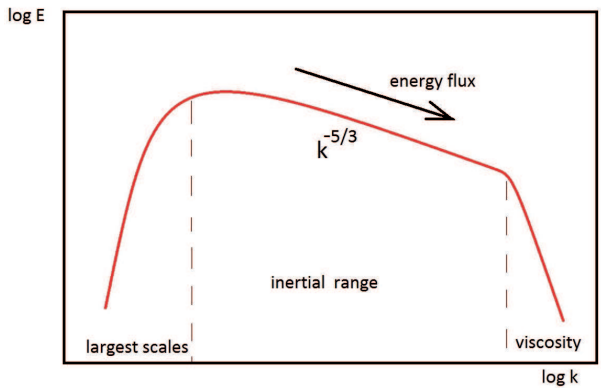}   
  \vspace*{-0.6cm}
\caption{Kolmogorov-Richardson spectrum. The energy is pumped into
the system via the largest-scale eddies, then it passes through the
scales of inertial range and dissipates at viscous scale. The
Kolmogorov's theory implies that the energy dissipation (and energy
flux) is independent of the Reynolds number. }
\end{figure}

This picture has inspired Kolmogorov for his K41 theory; the
independence of energy flux on the scale (as viscosity tends to
zero) is the central point of the theory. But the idea of the
cascade of breaking vortices, though produces phenomenological base
for K41, is not necessary for the rest of the theory, which is based
on dimensional considerations.

Despite its illustrativeness and historical role, the conception of
breaking vortices faces some problems. First, how exactly does
breaking of a vortex happen? To break, a vortex should bend  to make
an '8', and reconnect the vortex lines (Fig.3). However, the process
of reconnection is governed by viscosity $\nu$, and the
characteristic time of reconnection is $\sim l^2 /\nu$, where $l$ is
the characteristic width of the reconnecting region. Thus, in order
this time to be much smaller than the lifetime of the whole eddy
(roughly speaking, to exclude the dissociation of the whole eddy
before it reconnects many times, making a cascade), one has to
assume that the reconnection region is very thin; as $\nu \to 0$,
this implies presence of singularities at each reconnection. The
existence of finite-time singularity in solutions of the
Navier-Stokes (NS) and Euler ($\nu =0$) equations, which govern a
hydrodynamic flow, is an open question and a subject of many
investigations. If exist, the singularities could themselves produce
the observed statistical properties of turbulent flows, so the
cascade conception would not be needed any more. Anyway, the
existence of multiple vortex breaking depends on whether
singularities of such a special type exist in a flow or not.

\begin{figure}[h]
 \hspace*{-0.5cm}
\includegraphics[width=8cm]{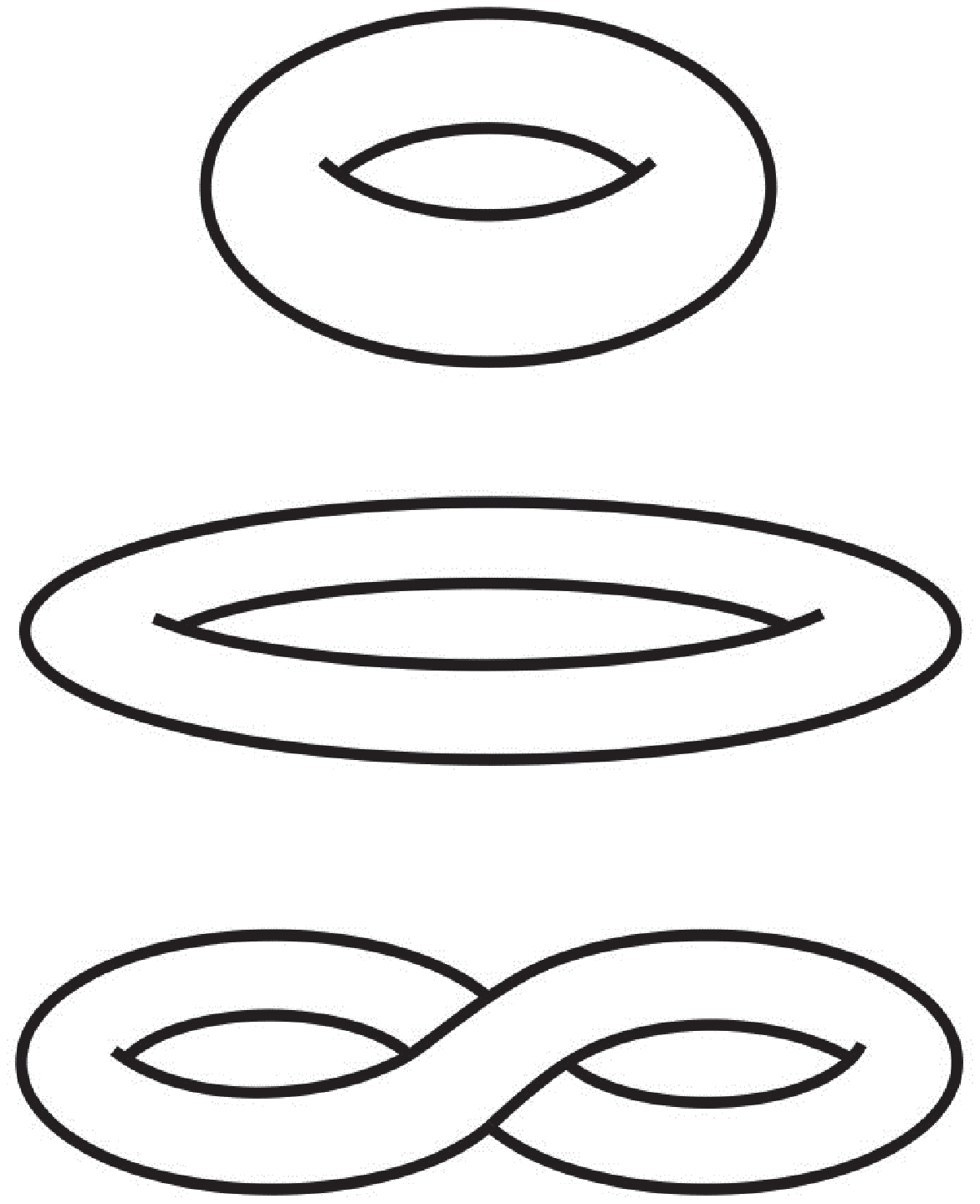}
  \vspace*{-1.6cm}  \hspace*{-0.5cm}

\includegraphics[width=8cm]{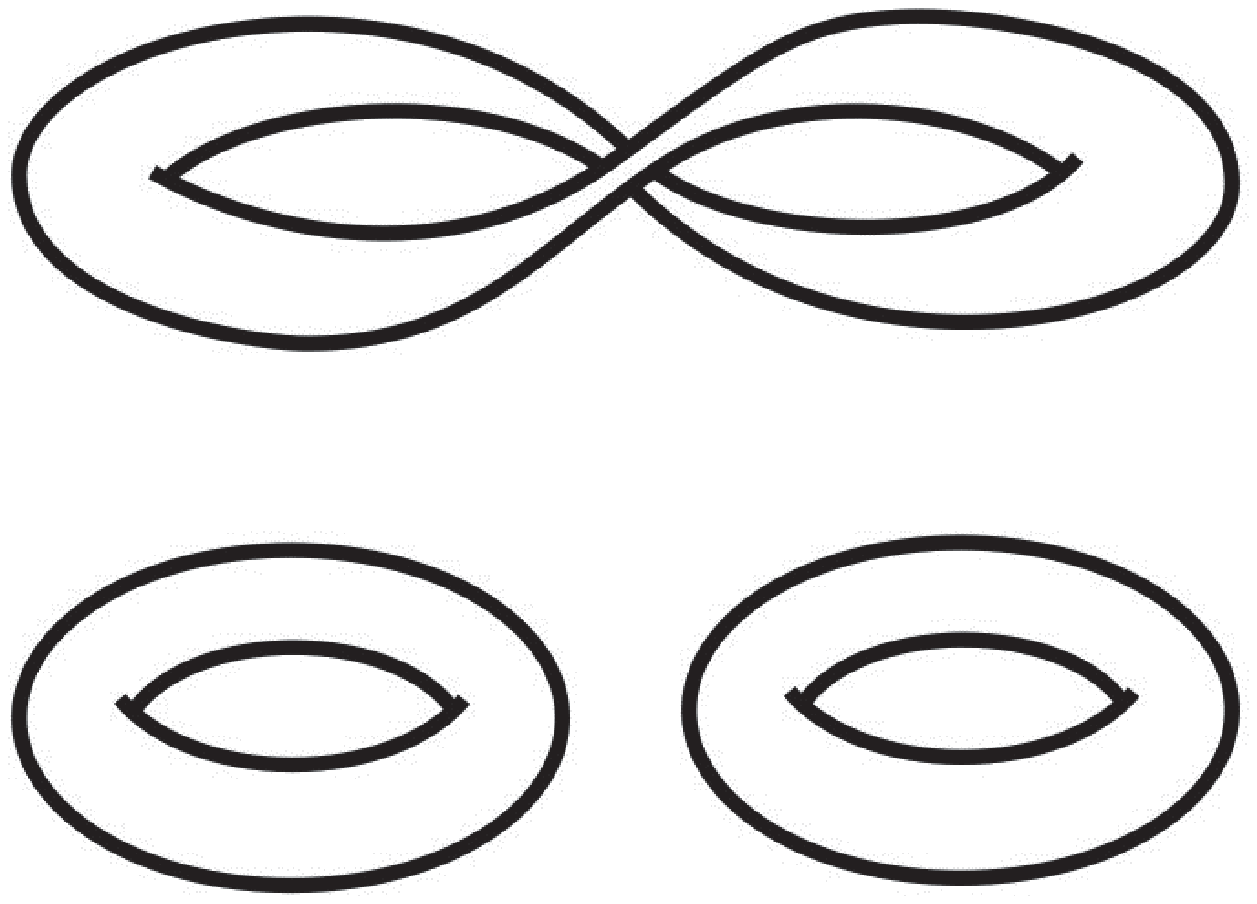}
  \vspace*{-1.2cm}
\caption{Breaking of a vortex. In this figure, time evolution of a
typical vortex in the cascade model is shown schematically. Vortex
lines are lying on the tori. To reconnect, the vortex has to become
extremely thin in the vicinity of the interception. In non-viscous
limit, singularity is needed to make the reconnection. }
\end{figure}

Second, many authors have reported observations  of long-living
'coherent structures' -- vortex filaments -- in experiments and in
direct numerical simulations (DNS) (see,e.g., \cite{Frisch208vnizu
or others?}). Recent investigations \cite{Farge}  have shown that
these filaments contain the most part of helicity of the flow, and
they are responsible for the observed energy spectrum corresponding
to the Kolmogorov's 5/3 law (Fig. 2). On the other hand, the
lifetime of these structures exceeds many times the largest eddy
turnover time; thus, their existence contradicts to the cascade
conception.

If not cascade, what physical mechanism could provide the energy
flux from larger to smaller scales? In the recent paper
\cite{PRE13}, we have studied the solutions of the  NS equation in
the regions of high vorticity. We have shown that motion of
particles in these regions is not just random: large-scale
perturbations cause a systematic exponential stretching combined
with random rotations. This stretching can be interpreted in terms
of evolution of vortex filaments; moreover, it is just the process
that may provide a physical picture of what happens in a turbulent
flow. It allows to explain the energy flux from larger to smaller
scales and to get scaling properties of statistical characteristics
without any assumptions of finite-time singularities. It also may
clarify the origin of the main assumption of the Kolmogorov's
theory: the energy dissipation independence of viscosity as $\nu \to
0$. Also the inequality of longitudinal and transverse statistics
may be understood from this new point of view.

In what follows we first describe the main features of the observed
solutions of the NS and Euler equations. Then we proceed to the
statistical description and discuss the connection between the
conception of stretching vortex filaments and the multifractal model
(MF).

As it is shown in \cite{PRE13}, the long-time asymptote of solutions
of the NS equation in the high vorticity regions can be described as
random rotation and systematic exponential stretching, which
corresponds to formation of a vortex filament. This stretching leads
to an exponential growth of vorticity: in some special rotating
coordinates, $(x,y,z)$,
$$
\omega \simeq \omega_z = e^{c
t} f\left(  e^{-a
t} x
\right)
$$
Here $a<0,c>0$ 
are parameters (Lyapunov
exponents) determined by  the properties of large-scale pulsations,
and $f$ is a function determined by boundary conditions near the
forming filament. Independently of the particular choice of $f$, the
solution 'shifts' all the initial ripples to the center of the
filament (Fig.4).
\begin{figure}[h]
\vspace*{-0.4cm} \hspace*{-0.5cm}
\includegraphics[width=8cm]{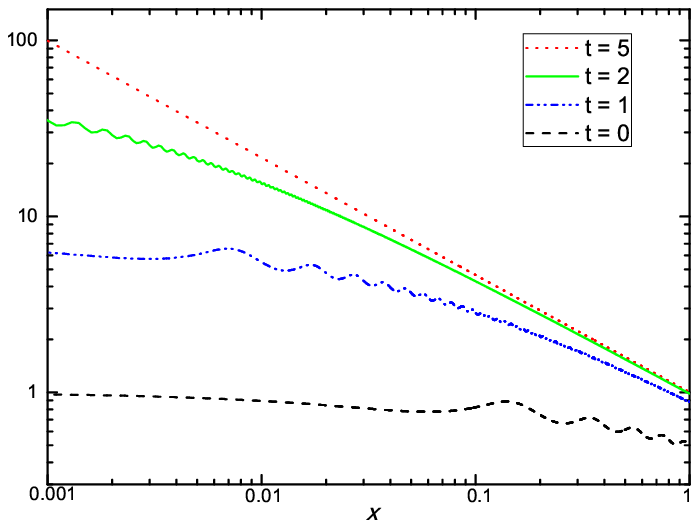}   
  \vspace*{-0.6cm}
\caption{Evolution of vorticity distribution in one particular
case:\\
$a=-3, c=2$, 
$\omega(0,x)=1/\left( 1+ \left[ x+0.1 \sin (10 \pi x) \right]^{2/3}
\right)$, $\omega(t,1)=1/ \left( e^{-2t}+ \left[ 1+0.1 e^{-3t} \sin
(10 \pi e^{3t}) \right]^{2/3} \right) $.
 One can see that the range of strong oscillations drifts to
smaller $x$, while inside the 'inertial range' the fluctuations
become negligible and the power law dominates. }
\end{figure}
For any reasonable boundary condition this results in formation of a
stationary (in the rotating frame!) power-law vorticity profile
$$
\omega (t,x) \sim x^{c/a} \ , 
$$
 while the exponentially
narrowing inner part remains non-stationary, and vorticity grows
exponentially inside the region. This is illustrated in Fig.5. Thus,
formation of vortex filaments provides scaling invariant velocity
and vorticity dependencies without appearance of a finite-time
singularity. In the case of small but non-zero viscosity, the growth
of vorticity stops as the contraction reaches the viscous scale. If
there is no viscosity, an infinite-time singularity appears in the
center of the filament, instead.
\begin{figure}[h]
\vspace*{-0.4cm} \hspace*{-0.5cm}
\includegraphics[width=8cm]{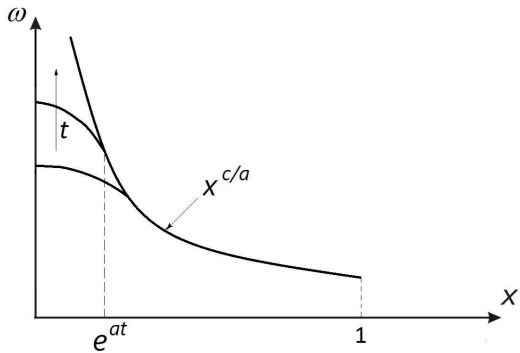}   
  \vspace*{-0.6cm}
\caption{Illustration of  
vorticity distribution as a function of 
time inside a vortex filament: the most part of the filament has
stationary power-law vorticity dependence, while the exponentially
shrinking core region remains non-stationary and provides the
exponential growth of the central vorticity. In the non-viscous
limit, after the infinite time this would become a singularity.}
\end{figure}

However, for any finite time $t$, the velocity and vorticity
distributions are smooth. The narrow non-stationary region in the
center of the filament, $x<x_0(t)=e^{a t}$ provides some non-viscous
analog, or substitute, of  viscosity: at $x>x_0(t)$, the non-viscous
solution does not differ from the 'viscous' one; the scale $x_0(t)$
thus has the meaning of the inner boundary of the inertial range.
Although the non-viscous equation does not 'smoothen' the initial
perturbations, it  transports them to smaller and smaller scales and
multiplies by a decreasing term. So, the solutions of Euler and NS
equations behave similarly in the limit $t\to \infty$, $\nu \to 0$.

 In Kolmogorov's theory, one assumes some definite order of taking
 limits: first $t\to \infty$, then $\nu \to 0$ and eventually $l\to
 0$.  In the model of contracting and stretching filament, one can
 get the same results with different order of limits: first,
 viscosity can be driven to zero, $\nu \to 0$; then $t\to \infty$ and
 finally, $l\to0$, or $t\to \infty$ and $l\to 0$ together, but $l>x_0(t)$.
 Thus, the model is less dependent on viscosity. This simplifies
 understanding of the main result of the Kolmogorov's theory even
 without its main assumption: even if $\nu=0$, and there is no dissipation
 at all,  the energy flux from
 larger to smaller scales remains constant.  Energy just continues
 to pass to smaller and smaller scales infinitely.

Quantitative description of turbulence implies the statistical
approach. Statistical properties of a flow can be described by
various correlation functions of velocity (and other related
quantities). The simplest  and the most widely used correlators are
velocity structure functions.
%
One distinguishes
longitudinal and transverse structure functions; they are defined by
$$  
S_n^{\parallel}(l)=\left< \left| \Delta {\bf v}\cdot \frac{\bf
l}{l}\right| ^n\right> \ , \quad
S_n^{\perp}(l)=\left< \left| \Delta {\bf v} \times \frac{\bf
l}{l}\right| ^n\right>
$$  
Here
$\Delta {\bf v} = {\bf v}({\bf r}+{\bf l}) -{\bf v}({\bf r}) $ 
 is velocity difference between two near points separated by ${\bf l}$,
 and the average is taken over all pairs of points separated by given $l$.

 Experiments and DNS show that inside the inertial range the structure functions
 obey scaling laws,
  $$ S_n^{\perp}(l)\propto l^{\zeta_n^{\perp}} \ ,
  \quad S_n^{\parallel}(l)\propto l^{\zeta_n^{\parallel}} \ .$$
  The scaling exponents $\zeta_n^{\perp}$, $\zeta_n^{\parallel}$  are believed
  to be independent on conditions of the experiment. They are intermittent, i.e.,
  $\zeta_n^{\perp}/n$ and $\zeta_n^{\parallel}/n$ are decreasing functions of~$n$.

 The most successful, and nowadays conventional, way to
interpret these properties is the multifractal (MF) approach
introduced in \cite{ParisiFrisch} . This is a generalization of
Kolmogorov's K41
theory and it allows us to express different 
intermittent  scaling characteristics (probability densities and
correlators of vorticities, accelerations, dissipation etc.) by
means of one function $D(h)$. This function has the meaning of a
fractal dimension of  a set ('$h$-class') near which local scale
invariance holds with the scaling exponent $h$ ($\delta v \propto
l^h$).

However, the MF model is based on dimensional consideration. Its
contemporary (probabilistic) formulation does not allow  us to
understand what happens in a turbulent flow; the structures (or the
solutions to the Navier-Stokes equation) that are responsible for
the observed intermittent properties remain unknown.
%
%
Intermittent behavior of the scaling exponents was obtained in
\cite{Lvov2001}  within the framework of the Sabra shell model,
which is a simplification of the NS equation in wave-vector
representation.
However, the relation of the results to solutions of the NS equation
remained unclear.

The idea of stretching vortices may help to fill this gap. It does
not contradict to the MF model; moreover, it gives the notion of
objects that contribute to correlations of each order, and to $D(h)$
(making different 'h-classes' responsible for different scaling
exponents).
As we will see below, it may also help to expand the
MF-model's validity.

 The question whether the longitudinal and transverse scaling exponents
 coincide in isotropic and homogeneous turbulence
  or not, is open. There is an exact statement that
  $\zeta_n^{\parallel}=\zeta_n^{\perp}$  for $n=2$
  and 3 \cite{Frisch}.
 On the other hand, modern
experiments \cite{Zhou,Shen} and numerical simulations
\cite{Chen,Benzi
} show 
noticeable difference between $\zeta_n^{\parallel}$ and
$\zeta_n^{\perp}$ at higher $n$. But the proponents of the equality
argue that the difference may result from finite Reynolds number
effects \cite{He,Hill
} or anisotropy \cite{BifProc}. Up to
now, these have been the only explanations to the observed
difference.

Before we continue to consider  the problem, we note that isotropic
and homogeneous medium does not necessarily imply
$\zeta_n^{\perp}=\zeta_n^{\parallel}$. For a simple illustration,
consider a gas of tops, each rotating around its own axis, the axes
directions  distributed randomly. Although locally a strong
asymmetry could be found near each top, the whole picture remains
isotropic. Vortex filaments might provide an analogous situation in
real turbulence.

In terms of the MF model, two different sets  of scaling exponents
($\zeta_n^{\parallel}$ and $\zeta_n^{\perp}$) correspond to two
different functions $D(h)$ \cite{25years}. But the MF model is  a
dimensional theory, it does not distinguish longitudinal and
transverse
structure functions. On the other hand, %
the conception of stretching vortices implies that %
scaling exponents of different sorts and orders are contributed by
different vortices. One can predict 
the shape of 'extreme' vortex 
that provides the
largest-order ($n\to \infty$) exponents: it corresponds to $h=0$ in
the MF model, which means that $\delta v(l) \sim l^0$
According to our conception, stretching is the main process  that
manages the vortices, and stretching along one axis is  much
stronger than that in two other directions. So, the expected
'extreme' vortex filament must have an axial symmetry and the
velocity profile:
 \begin{equation}\label{extrim}
{\bf v} = [{\bf e}_z, {\bf r}/r]   \ , 
\end{equation}
We see that rotational  velocity is independent of the distance $r$
to the filament's axis $z$. The pressure diverges logarithmically in
such a construction, which means that any smoother velocity profile
(corresponding to infinitesimal values of $h$) is possible.
(Negative values of $h$ are forbidden by infinite velocities and
infinite pressure.)

Calculation of the correlators 
corresponding to the spatial distribution (\ref{extrim}) gives under
the limit $n\to \infty$:
\begin{equation}
\label{extrim2}
  \langle \left| \Delta v \times {\bf l}/l \right| ^n  \rangle \propto
  \frac{2^n}{n} \, l^2\,,\quad
  \langle \left| \Delta v \cdot {\bf l}/l \right| ^n \rangle \propto
   n^{-5/2} l^2
\end{equation}
The proportionality to $l^2$ is caused by the axial geometry of the
filament (integrating $r dr$). This corresponds to the definition of
$D$ given in, e.g., \cite{Frisch}: the probability that at least one
of a pair of points would get inside the filament is proportional
to $l^2$. Thus, we get
  \begin{equation}
 \label{condperp}
D^{\perp}(0)=1 \ , \quad   \zeta_n^{\perp} =2
\end{equation}
However, while the contribution of the filament (\ref{extrim2}) to
the transverse structure functions increases and becomes
determinative  as $n\to \infty$, its contribution to
$\zeta_n^{\parallel}$ decreases because of the diminishing
pre-exponential multiplier. Indeed, the 'extremely cylindric' vortex
does not produce longitudinal velocity differences.

There must be other solutions to the Euler equation to determine the
behavior of $D^{\parallel}(h)$ for small $h$ and, equivalently, to
make the most contribution to $S_n^{\parallel}$ for large $n$.
Curvature of the 'extreme cylinder'  provides a subleading term
differing by  $l/L$ where $L$ is the curvature radius of the
filament; for longitudinal structure functions, this subleading term
becomes the main term. Thus, the needed
solutions correspond to 'strongly curved' extreme filament. To
satisfy $h=0$, velocity must be independent of $r$. It may, for
example, take the form
 ${\bf v} = (v_r(\theta), v_{\theta}(\theta), 0)$ in spherical coordinates.
  Such a solution  exists but it cannot be written analytically.
   Since $\delta v \sim l^0$ in the
case, and averaging includes $r^2 dr$, the correlator is
proportional to $l^3$. This corresponds to
 \begin{equation}
 \label{condpar}
D^{\parallel}(0)=0 \ , \quad \zeta_n^{\parallel}=3
\end{equation}
The difference between the boundary values (\ref{condperp}) and
(\ref{condpar}) determines  the difference between the functions
$\zeta_n^{\perp}$ and $\zeta_n^{\parallel}$.

By construction, the function $D(h)$ must be growing and concave
function of $h$ in the range $0\le h \le h_{min}$.  The universal
condition $\zeta_0^{\perp,\parallel} =0$ gives the  restriction
$\max \limits_h {D(h)}=3 $. One more fundamental restriction for
$D(h)$ comes from the Kolmogorov's 4/5 law:
$\zeta_3^{\perp,\parallel}=1$.
 Now as we have the boundary values $D^{\perp}(0)$ and
 $D^{\parallel}(0)$,  we can estimate the rest of $D(h)$ just
 assuming it to be a parabola. Doing this leads to
\begin{equation}   \label{result}
\begin{array}{ll}
\zeta_n^{\parallel} = & \left\{
\begin{array}{lr}
 0.367 n - 1.12\cdot 10^{-2} n^2 \ , &   
\quad n \le  16.3  \ ; \\
3 \ , & \quad n>16.3 \ .
\end{array}     \right.
\\ 
\zeta_n^{\perp} = & \left\{
\begin{array}{lr}
 0.391 n - 1.91 \cdot 10^{-2} n^2 \ , &
\quad n \le  10.2  \ ; \\
2 \ , & \quad n>10.2 \ .
\end{array}     \right.
\end{array}
\end{equation}
In Fig. 6 we compare this theoretical prediction with  the DNS data
by \cite{Benzi} and \cite{Gotoh}. We see that the result of our
simple model is very close to the experimental results and lies
inside the error bars of the experiments.

\begin{figure}
\vspace*{-0.4cm} \hspace*{-0.5cm}
\includegraphics[width=8cm]{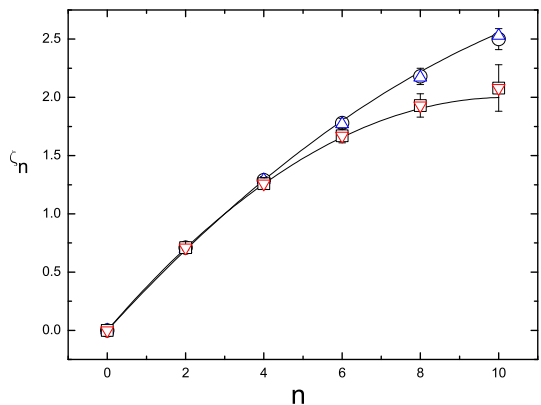}   
  \vspace*{-0.6cm}
\caption{Euler longitudinal (upper branch) and transverse (lower
branch) scaling exponents:  the results of DNS from \cite{Benzi}
($\bigcirc$,   $\square$) and \cite{Gotoh}  ($\bigtriangleup$.
$\bigtriangledown$), and the prediction of the theory (\ref{result})
(lines). }
\end{figure}

Our simple model has one difficulty: the two parabolas plotted  in
Fig.~6 coincide at $n=0$ and $n=3$, hence they do not coincide at
$n=2$. This contradicts to the exact theoretical statement that
$\zeta_2^{\parallel} = \zeta_2^{\perp}$. However, this difficulty is
caused by postulating the simplest parabolic shape for $D(h)$. It
can easily be eliminated by 
assuming
(at least one of the two) $D(h)$ to be a cubic polynomial.
 Because of very small divergence ($1.6 \cdot 10^{-2}$) between
 $\zeta_2^{\parallel}$ and
$\zeta_2^{\perp}$ in Fig.~6, the coefficient by the eldest order
must be very small ( $   \le 10^{-4} $).
 It would change very slightly (unnoticeable for an eye)
the lines presented in Fig.~6. The only thing it may  change
significantly is the rate of approaching the constant at large (but
still intermediate) $n$ (from 10 to 15, approximately). But this
range of $n$ is, anyway, badly described by the lowest-order
polynomials: adding more degrees of freedom with very small
coefficients, though unimportant for smaller $n$, would change the
solutions for these $n$. However, the changes cannot be very big,
since the exponents are still restricted by the values 2 and 3,
respectively.

This saturation of the scaling exponents as $n\to\infty$  is an
exact prediction of our model. This agrees well with
\cite{Yakhot2001} where the possibility of saturation of $\zeta_n$
was also noticed. Saturation seems to be a typical feature of
turbulent scaling exponents: the same property is demonstrated in
the passive scalar theory \cite{Falkovich}.


One more comment is that, knowing $D(h)$, one can use the MF model
to calculate, e.g., the probability distribution function of
velocity gradients or accelerations. In \cite{25years} it is shown
that once $D(h)$ fits $\zeta_n$ well, it would also fit well the
other quantities.

\vspace{0.3cm}

So, the conception of stretching vortices agrees well with the MF
model and the observed statistical properties of hydrodynamic
turbulence. The change of conception itself, without any additional
assumptions, allows to realize what are the 'extreme' structures
contributing to the highest-order correlators: the result of vortex
breaking is indefinite, whereas the result of vortex stretching is
predictable. The conception of stretching vortices thus immediately
leads to verifiable predictions: it implies inequality of
longitudinal and transverse velocity scaling exponents and their
saturation at the values 3 and 2, correspondingly.



We thank Prof.A.V. Gurevich for his kind interest to our work. The
work was partially supported by RAS Programm 19 $\Pi$ "Fundamental
Problems of Nonlinear Dynamics in mathematics and physics"

\end{document}